# Single-Shot Local Measurement of Terahertz Correlated Second Harmonic Generation in Laser Air Plasma Filaments

MERVIN LIM PAC CHONG,‡ KAREEM J. GARRIGA FRANCIS,‡ YIWEN E, AND XI-CHENG ZHANG*

‡ *Mervin Lim Pac Chong and Kareem J. Garriga Francis contributed equally*
*Corresponding author: xi-cheng.zhang@rochester.edu*

**We present a single-shot detection method of terahertz-correlated second harmonic generation in plasma-based sources by directly mixing an optical probe into femtosecond laser-induced plasma filaments in air. The single-shot second harmonic trace is obtained by measuring second harmonic generation on a conventional CCD with a spatio-temporally distorted probe beam. The system shows a spectrometer resolution of 22 fs/pixel on the CCD and a true resolution on the order of the probe pulse duration. With considerable THz peak electric field strengths, this formalism can open the door to single-shot THz detection without bandwidth limitations.**

In recent years, the development of large-scale laser facilities has led to major contributions in Terahertz (THz) and plasma sciences. Recently, >1 J lasers showcased record-setting optical-to-THz energy conversion in Lithium Niobate exceeding 1.2%[1]. Similar facilities were also used to produce large THz peak electric field strengths above MV/cm in plasma[2]. The traditional method to fully characterize the THz emission from sources has been through electro-optic sampling (EOS)[3–6]; but, with increasing THz electric field strength, electro-optic (EO) crystal over-rotation poses an issue in accurately representing a signal with minimal distortion. The crystal structure also limits the spectral detection bandwidth.

Terahertz Field-Induced Second Harmonic (TFISH) generation offers a platform for broad spectral detection range. It mixes optical photons with THz photons in a third-order nonlinear process to produce a second harmonic (SH) of the fundamental optical photons[7]. The TFISH signal can also be mixed with a reference SH signal to yield a coherent signal through homodyne detection.

In this letter, a potential method for single-shot detection of a broadband THz pulse is presented by directly mixing an optical probe onto a two-color femtosecond laser-induced air plasma. We designed a noncollinear TFISH generation by mixing a probe beam onto the plasma at 40° incidence with respect to the pump propagation axis. In previous results, we showed that the system provides an SH signal that is correlated to the THz field through the TFISH model[8], and the temporal and polarization dynamics were studied in ref. [9]. Despite other interpretations of SH in plasma[10-13], the temporally gated signals found in this study and the high conversion efficiency are new. It is currently not definitive that the appearance of the strong SH is explicitly field induced in the traditional sense. Since many linear and nonlinear processes occur in the creation of a plasma filament, theoretical analysis of the system is complicated. Work to determine the authenticity of the signals as THz field induced or a consequence of the same physics is ongoing. We continue to use TFISH in this paper to easily present the data.

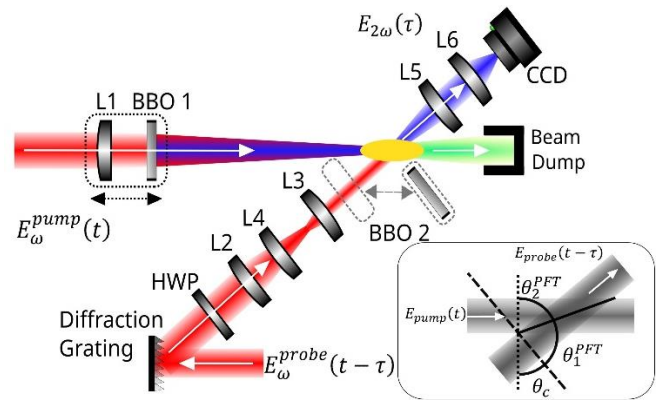

Fig. 1: Experimental setup. The pulse-front tilt (PFT) is imaged onto the interaction region by a Keplerian telescope. A CCD detects the SH generated, which is isolated with a dichroic mirror (Thorlabs DMLP490L) and two bandpass filters (Thorlabs FBH400-40). BBO 1 and BBO 2 are β-BBO crystals. L1-L6 are lenses. HWP is a half-wave plate. Inset shows the angles at the interaction region: $\theta_C$ is the pump-probe angle and $\theta_i^{PFT}$ is the PFT angle where $i$ is 1 or 2. The dashed and dotted lines are the phase fronts of the probe and pump beams, respectively, and the solid line represents the probe PFT.

As shown in Fig. 1, the noncollinear TFISH system consists of a two-color air-plasma THz source where the probe beam intersects the plasma at 40° incidence. A Coherent Astrella amplified laser system operating at 6.5 W, 1 kHz repetition rate, and near 100 fs is used. The laser fundamental wavelength is 800 nm (ω) with a 12 mm at $1/e^2$ initial beam spot size. The pump and probe paths are 5.2 W and 1.3 W, respectively. A 100μm-thick type I β-Barium Borate (BBO 1) crystal is placed after the 300 mm pump lens (L1) to maximize the THz field strength.

An 800 lp/mm grating is used to produce a 40° pulse-front tilt (PFT) along the intensity-front of the probe with respect to its wavefront[14], and thus encode temporal information onto spatial coordinates. This configuration is hereby referred to as a single-shot/real-time PFT system. The PFT is imaged from the grating to the lens L1's focal region with a Keplerian telescope consisting of a pair of cylindrical lenses with focal lengths 300 mm (L2) and 100 mm (L3). This is considered the "grating-imaging system". Since the PFT probe is effectively collimated in the interaction region between the probe and the plasma, a third cylindrical 200 mm lens (L4) is used to focus the beam orthogonally (vertical axis) to the induced PFT. The result is a higher intensity along a line in the focal region. The interaction is imaged to a CCD by a Keplerian telescope using spherical lenses. Filters and dichroic mirrors are used to discriminate the TFISH signal from the fundamental probe.

To achieve coherent detection, a second 100μm thick type I β-Barium Borate, labeled as BBO 2 in Fig. 1, is used in the probe beam path to produce a reference wave. When the polarization of the SH wave induced by BBO 2 has components along the same axis as the polarization of the TFISH signal, an interferometric process leads to a coherent homodyne detection of THz field. Hence, a half-wave plate (HWP) and the BBO 2 orientation are used to adjust the former. A delay line in the probe beam path aids in spatio-temporal overlap.

The multi-shot behavior of the noncollinear system and the TFISH signals retrieved at pump-probe spatio-temporal overlap have been shown in ref. [8]. Due to the nature of a plasma-filament, there exists a localized TFISH signal maxima because of the varying electron number density along the propagation axis, which is found by moving L1 and BBO 1 together through the probe beam focus[8].

The noncollinear system can be modeled using the TFISH process to retrieve a signal:
$$S_{2\omega} \propto \int_{-\infty}^{\infty} |\chi^{(3)}(2\omega, \omega, \omega, \omega_{THz}) E_\omega^2(t - \tau - \alpha_{PFT} x) E_{THz}^*(t)|^2 \partial t, \quad (1)$$
where $S_{2\omega}$ is the signal detected by the PMT, $\chi^{(3)}(2\omega, \omega, \omega, \omega_{THz})$ is the third-order susceptibility tensor of air within the plasma, $E_\omega(t \pm \tau - \alpha_{PFT} x)$ is the electric field of the probe beam delayed by a temporal factor $\tau$ and influenced by PFT with $\alpha_{PFT}$ in s/m, $x$ is the transverse coordinate, and $E_{THz}(t)$ is the THz field confined to the plasma filament. The probe beam is incident onto the grating at 0° leading to an initial diffraction at 40°. The PFT at the grating is directly imaged to the interaction region. The linear propagation from the grating to the CCD is modeled using Kostenbauder matrices[15, 16] (K-matrices), yielding a total induced PFT angle of 70° at the interaction region with the following 4x4 K-matrix:

$$K_{int} = \begin{bmatrix} \frac{f_3}{f_2\sigma} & 0 & 0 & 0 \\ 0 & \frac{f_2\sigma}{f_3} & 0 & -\frac{f_2\lambda\xi}{f_3 c} \\ -\frac{\xi}{c\sigma} & 0 & 1 & 0 \\ 0 & 0 & 0 & 1 \end{bmatrix} = \begin{bmatrix} A & B & 0 & E \\ C & D & 0 & F \\ G & H & 1 & I \\ 0 & 0 & 0 & 1 \end{bmatrix},$$

where $f_2$ and $f_3$ are the focal lengths of $L_2$ and $L_3$, respectively, $\sigma$ is the angular grating magnification, $\lambda$ is the probe beam fundamental center wavelength, $\xi$ is the grating angular dispersion parameter, and c is the speed of light in air[17]. Note that L4 is omitted since it has no effect in the angle of the PFT. Parameters A, B, C, and D of the K-matrix form the well-known ABCD matrix, with a temporal counterpart in the bottom right quadrant with parameter I being the group delay dispersion. The spatio-temporally coupled terms E, F, G, and H denote spatial chirp, angular dispersion, PFT, and time vs angle, respectively[15,16]. A spatio-temporal Gaussian pulse yields a reduced Q-Matrix at the interaction region given by[16]:

$$Q = \frac{\begin{bmatrix} A & 0 \\ G & 1 \end{bmatrix} \cdot Q_{in} + \begin{bmatrix} B & \frac{E}{\lambda} \\ H & \frac{I}{\lambda} \end{bmatrix}}{\begin{bmatrix} C & 0 \\ 0 & 0 \end{bmatrix} \cdot Q_{in} + \begin{bmatrix} D & \frac{F}{\lambda} \\ 0 & 1 \end{bmatrix}} = j\frac{\lambda}{\pi}\begin{bmatrix} Q_{xx} & Q_{xt} \\ -Q_{xt} & Q_{tt} \end{bmatrix}^{-1}.$$

The output field becomes $E(x, z, t) = e^{[x^2 Q_{xx} + 2xt Q_{xt} - t^2 Q_{tt}]}$, where $Q_{xx}$ describes the spatial Gaussian parameters, $Q_{xt}$ is the coupling term, and $Q_{tt}$ describes the temporal width of the pulse. The input Q-matrix $Q_{in}$ (before the grating) describes the initial probe, with $Q_{xt} = 0$. The PFT is extracted from Q by $\alpha_{PFT} = \frac{\Re\{Q_{xt}\}}{\Re\{Q_{tt}\}}$, and described by the angle $\theta_{PFT} = tan^{-1}(\alpha_{PFT} c)$ between the beam's intensity front and wavefront[14]. The extension of the K- and Q-matrices to the CCD plane from the grating plane is simulated and carried out in MATLAB.

The noncollinear geometry leads to an effective PFT angle $\theta_2^{PFT} = 180° - \theta_1^{PFT} - \theta_C$, where $\theta_2^{PFT}$ is the PFT angle in the interaction region, $\theta_1^{PFT}$ is the PFT angle imaged with the first telescope, and $\theta_C$ is the angle between the pump and probe. In the interaction region, the beam is an ellipse with horizontal and vertical waist radii of 1 mm and 9 μm, respectively. Careful alignment is vital for minimum TFISH trace size during the spatio-temporal overlap.

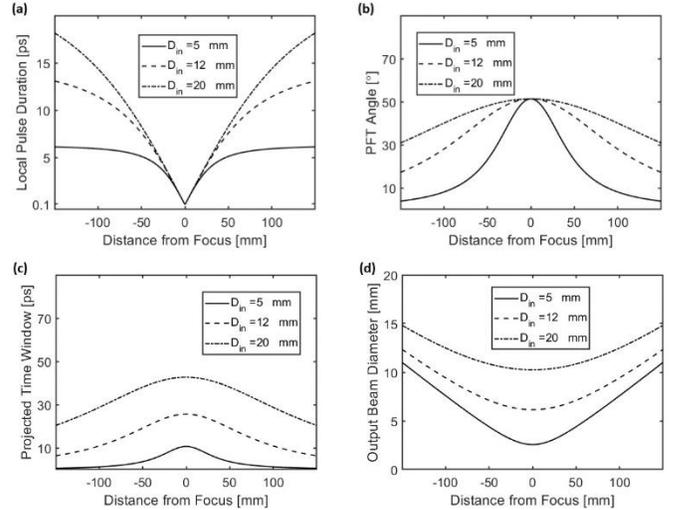

Fig. 2: Simulation results of a K-Matrix solver for system optimization using three initial probe beam diameters. (a) Beam pulse duration lengthening due to angular dispersion as a function of distance from the CCD (lens focal plane). (b) Maximum achievable PFT vs distance from the CCD. (c) Maximum achievable time window vs distance from CCD. (d) Output beam diameter vs distance from the CCD.

Figure 2 shows the simulations of the PFT system using a K-matrix solver from the grating to the CCD for initial probe beam diameters $D_{in}$ = 5 mm, 12 mm, and 20 mm. The simulation accounts for the doubled PFT magnification by the Keplerian telescope after the interaction region. Although the PFT angle reduces from 70° to 50°, the setup becomes physically easier. From Fig. 2(a), the grating image plane is the only point where the probe pulse-duration is preserved, and spatio-temporal overlap is undistorted—yielding the maximum possible PFT angle for a given system as shown in Fig.

2(b). Moving away from the focus leads to a larger pulse duration and thus, loss of information according to Eq. 1. In Fig. 2(c), as the initial diameter is increased, the tolerance in localizing the maximum PFT angle and time window are relaxed at the trade-off of pulse duration as seen in Fig. 2(a). This is due to the elimination of spatial chirp and group delay dispersion at the overlapping point, which has a 0.6 mm theoretical tolerance for an 800 lines/mm grating.

Figure 2(d) shows that as the beam diameter onto the grating is increased, the divergence from angular dispersion is relaxed. In our experiment, a 12 mm diameter beam was chosen because it yielded good experimental tolerances without needing large aperture optics.

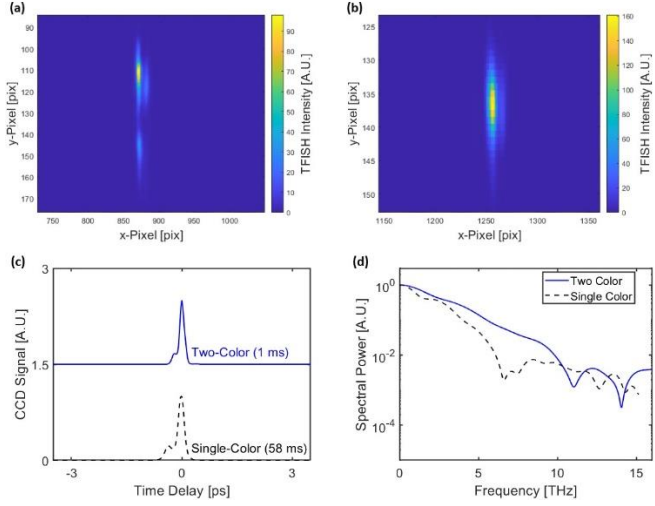

Fig. 3: Real-time detection results. (a) and (b) are the real-time SH image gathered at the CCD for single- and two-color pumped signal, respectively. (c) Temporal TFISH signal for two- (solid line) and single-color (dashed line) pumps resulting from integrating over the row (y-) pixels. The waveforms are flipped along the temporal axis to match previous results and vertically shifted for clarity. (d) Spectra corresponding to the TFISH signals.

The overlapped spatial region directly yields a temporal coordinate through expressions[6] $T_W = \frac{Dtan\theta_2^{PFT}}{c}$ and $\Delta t_R = \frac{\Delta x tan\theta_2^{PFT}}{c}$, representing the temporal window and device-limited resolution, respectively. Here, $T_W$ represents the spatio-temporally coupled total detection time window, D is the optical beam diameter, $\Delta t_R$ is the temporal resolution per CCD pixel, and $\Delta x$ is the CCD pixel width. When the interaction region is directly imaged to the CCD, a y-pixels integration leads to the temporal detection of the TFISH signal due to the linear relationship between $T_W$ and D.

In Fig. 3(a), the real-time trace of the TFISH signal for a single-color excitation pump plasma is achieved by omitting BBO 1. Because the single-color signal is weak, higher probe pulse energy (1.2 mJ at 1 kHz) and longer exposure time (60 ms) on the CCD are required to produce a visible trace. The result remains a real-time trace where no delay scan is needed and can be done in single-shot with a 9.3 mJ pulse energy. Decreasing the probe beam diameter for lower single-shot energy operation will incur a cost on the time window. Furthermore, increasing intensity and the trace visibility by reducing L3 inherently increases the PFT angle, lowering the temporal resolution. In Fig. 3(b), because the TFISH trace shown is for a two-color pump plasma, even a <130 µJ optical probe beam energy yields good visibility at an exposure time of 1 ms. This signal is effectively single-shot and theoretically requires no average.

To retrieve the temporal trace, we average over every row of pixels. The traces for a single-color and two-color source are shown in Fig. 3(c) at an exposure of 58 ms and 1 ms, respectively, and their corresponding power spectrum is shown in Fig. 3(d). The two-color trace is easily visible to the human eye. Figure 3 is background-free because the process, according to Eq. 1, is inherently so. Other secondary SH radiation processes may still lead to the appearance of DC components in Fig. 3(d).

The CCD used in the experiment (Imaging Source DMK 27BUP031) has a 2.2 µm pixel size. Experimentally, the PFT method shows a resolution of 22 fs/pix and is measured by the time difference ratio over the number of pixels the peak of a signal shifted between two different time delay positions. This spectrometer resolution is different from the real resolution. The latter is found through the convolution between the probe optical pulse and Gaussian spectrometer function with a width set to 22 fs, resulting in a resolution near the probe pulse duration. Since the PFT method encodes the temporal information directly onto the spatial width of the probe beam, a very large time window is accessible. In Fig. 3(c), despite the maximum achievable time window is within the theoretical expectation of 15 ps, the THz beam size limits it to 1 ps. While relatively short, this method can still be expanded to systems with larger THz beam diameter. For example, a 15 ps time window is possible with only a 2 mm diameter focused THz beam.

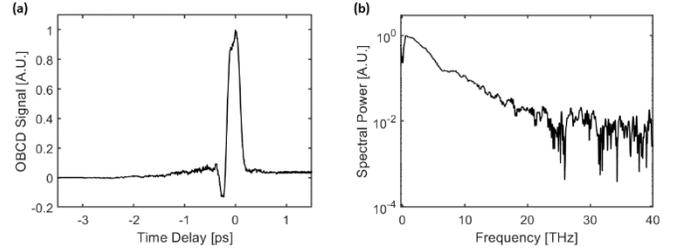

Fig. 4: (a) Single-shot OBCD signal from the PFT configuration gathered at 1 ms exposure time. (b) Corresponding Spectral power of (a) obtained through mathematical Fourier Transform.

The coherent signal for the two-color source is found through the interferometric mixing of the TFISH signal with a controlled SH. Using the BBO 2 crystal, the 2ω (400 nm) probe is termed a Local Field-Induced Second Harmonic (LFISH) generated wave. Considering that some of the ω probe beam propagates unconverted, the interferometric mixing of the two SH signals takes the form:

$$S_{2\omega} \propto |E_{2\omega}^{LFISH} + E_{2\omega}^{TFISH}|^2 . \qquad (2)$$

The above expression produces two incoherent intensity traces and a cross-correlated term known as the Optically Biased Coherent Detection (OBCD) term[18]. A background subtraction operation is required to extract the OBCD term. The OBCD trace for a two-color source is shown in Fig. 4(a) along with its Fourier transform in Fig 4(b). In this case, probe instabilities clearly outline the time window as limited by the width of the THz beam (since it is smaller than the horizontal width of the probe). The PFT coherent detection is done with a 1 ms exposure time, making it effectively single-shot for our 1 kHz repetition rate laser pulses.

The THz spectrum can still be limited by alignment constraints. The BBO 2 crystal has a small aperture and part of the PFT trace, and thus the probe spectrum, is clipped as it must be placed in a region of dominant spatial chirp to avoid crystal damage. When the PFT is established at the interaction region, the clipped trace results in an artificial lengthening of the probe pulse from the time-bandwidth product. Because the beam is spatio-temporally coupled, the signal becomes slightly distorted, and the maximum detection spectrum decreases proportionally to the increase in the probe pulse duration. Before the interaction region, the temporal information is not yet localized along the probe beam and, from angular dispersion, clipping the edges of the beam can lead to reduced information in the overall signal. Yet, the recovered spectrum still boasts detection capabilities beyond what is possible with single-shot EOS.

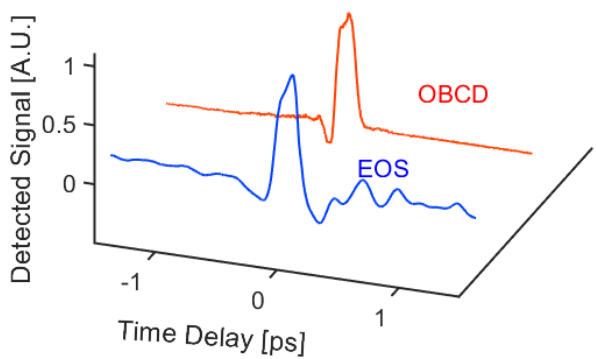

Fig. **5**: A comparison of a plasma source characterized by single-shot EOS (blue) and single-shot OBCD (red).

For completeness, the single-shot OBCD signal is further compared with a PFT-based single-shot EOS signal from setups like refs. [19, 20] in Fig. 5. In our case, the PFT for single-shot EOS is induced with a 1200 lp/mm grating and the detection crystal is a 1 mm thick ZnTe. The same plasma source and CCD are used in both cases. Table 1 compares the performance of both systems. With similar Signal-to-Noise Ratio (SNR) and Dynamic Range (DR), the tenfold improvement in detection bandwidth makes this work a better option, especially for higher spectral components analysis.

**Table 1** Comparison of system parameters obtained from Fig. 5 for a single-shot EOS and the single-shot OBCD (This Work) applied to the same plasma source.

| System | EOS | OBCD (This work) |
|---|---|---|
| SNR | ~200 | ~200 |
| DR | >10$^3$ | >10$^3$ |
| Bandwidth | 2.5 THz | >20 THz |
| Probe Energy Required | < 10 nJ | < 130 µJ |

In summary, we present local measurement of THz intensity profile and electric field profile inside a plasma by mixing a spatio-temporally coupled probe beam directly into the plasma. The high visibility resulting trace serves as a powerful diagnostic of THz radiation in plasma-based sources. A broad detection bandwidth is recovered for a two-color air-plasma THz source. Since single-shot EOS concept works here, the detection window and resolution can be improved with other spatio-temporal couplings. Angular dispersion without PFT can yield high resolution and high temporal window characterization of THz waves.

**Funding.** This work is supported by the Air Force Office of Scientific Research (FA9550-21-1-0389 and FA9550-21-1-0300) and National Science Foundation (Grant no. ECCS-2152081).

**Acknowledgments.** The authors would like to thank Shing Yiu (Steven) Fu for his input on the original system design, Jiacheng Zhao and Zhigong (Archie) Gao for their help with experiments that led up to this work.

**Disclosures.** The authors declare no conflicts of interest.

**Data availability.** Data underlying the results presented in this paper are not publicly available at this time but may be obtained from the authors upon reasonable request.